\documentclass[aps,pre,twocolumn,showpacs,superscriptaddress]{revtex4}

\usepackage[normalem]{ulem}
\usepackage{color}
\usepackage{epsfig}
\usepackage{amsfonts}

\begin{document}

\title{The origin of power-law distributions in deterministic 
walks: the influence of landscape geometry}

\author{M. C. Santos}
\affiliation{Departamento de F\'{\i}sica,
Universidade Federal do Paran\'a,
81531-990 Curitiba--PR, Brazil.}

\author{D. Boyer}
\affiliation{Departamento de Sistemas Complejos,
Instituto de F\'{\i}sica,
Universidade Nacional Aut\'onoma de M\'exico,
Apartado Postal 20-364, 01000 M\'exico D.F., M\'exico}

\author{O. Miramontes}
\affiliation{Departamento de Sistemas Complejos,
Instituto de F\'{\i}sica,
Universidade Nacional Aut\'onoma de M\'exico,
Apartado Postal 20-364, 01000 M\'exico D.F., M\'exico}

\author{G. M. Viswanathan}
\affiliation{Instituto de F\'{\i}sica,
Universidade Federal de Alagoas,
57072-970 Macei\'o--AL, Brazil}

\author{E. P. Raposo}
\affiliation{Laborat\'orio de F\'{\i}sica Te\'orica e Computacional,
Departamento de F\'{\i}sica,
Universidade Federal de Pernambuco,
50670-901 Recife--PE, Brazil}

\author{J. L. Mateos}
\affiliation{Departamento de Sistemas Complejos,
Instituto de F\'{\i}sica,
Universidade Nacional Aut\'onoma de M\'exico,
Apartado Postal 20-364, 01000 M\'exico D.F., M\'exico}

\author{M. G. E. da Luz}
\email{luz@fisica.ufpr.br}
\affiliation{Departamento de F\'{\i}sica,
Universidade Federal do Paran\'a,
81531-990 Curitiba--PR, Brazil.}


\begin{abstract}

We investigate the properties of a deterministic 
walk, whose locomotion rule is always to travel to the 
nearest site.
Initially the sites are randomly distributed in a closed
rectangular ($A/L \times L)$ landscape and, once reached, 
they become unavailable for future visits.
As expected, the walker step lengths present characteristic 
scales in one ($L \rightarrow 0$) and two ($A/L \sim L$) 
dimensions.  
However, we find scale invariance for an intermediate 
geometry, when the landscape is a thin strip-like region.
This result is induced geometrically by a dynamical
trapping mechanism, leading to a power law distribution for 
the step lengths.
The relevance of our findings in broader contexts 
-- of both deterministic and random walks -- is also briefly 
discussed.

\end{abstract}


\maketitle

\section{Introduction}

A large number of phenomena in physics, ecology, chemistry,
economics, etc~\cite{metzler1,zaslavsky1,peterson}, 
characterized by scale invariant distributions, are in many 
situations associated with L\'evy walks and L\'evy 
flights~\cite{shlesinger1,tsallis1,klafter}.  
Furthermore, when related to diffusion mechanisms, these types of 
systems present mean square displacements that, for large enough 
times~\cite{criterio}, scale as $t^\alpha$ with $\alpha > 1$.

In such contexts, some of the relevant challenges are to determine: 
(a) if there are global driving forces underlying the superdiffusive
features; 
(b) how they can emerge; and 
(c) how they are linked to other properties such as self-similarity, 
fractality, noise with $f^{-\beta}$ power spectrum and intermittent 
bursts behavior, ubiquitous in Nature~\cite{mandelbrot}.  
In fact, many studies address such general questions under different 
perspectives.  
For instance, one idea points to the concept of self-organized 
criticality~\cite{soc}.  
The so called spatiotemporal complex systems evolve through a series 
of avalanches towards critical states, which possess scale invariance 
and long range correlations.  
These hierarchical ``paths'' are unavoidable due to the character 
of the dynamics and are observed in many problems~\cite{paczuski}.  
For deterministic chaotic systems, on the other hand, the above 
queries may be associated either to dynamical fractional 
kinetics~\cite{zaslavsky1} or to phase space strange non-chaotic 
attractors and attracting sets of particular geometrical partitions 
(see~\cite{zaks} and refs. therein).

In the realm of stochastic processes, especially random walks, there
are different direct causes for superdiffusion and power-law tailed
decay.  
To exemplify just a few of them, we mention: 
(i) the evolution governed by fractional Fokker-Planck 
equations~\cite{metzler1}; 
(ii) dichotomic systems for which two time scales, microscopic and
macroscopic~\cite{west}, can be identified; 
(iii) the existence of correlations in the variance of the 
physically relevant quantities~\cite{variance}; and 
(iv) random multiplicative processes in the presence of a boundary 
constraint (a repealing barrier)~\cite{multiplicative}.
Moreover, it may happen that a L\'evy or power law distribution 
for the dynamical variables of a random process may be a natural 
way to lead to certain outcomes such as: 
the diversity of species in evolutionary ecology~\cite{evolutionary}; 
efficiency optimization in random search, e.g., animal 
foraging~\cite{foraging} in continuous Euclidean 
spaces~\cite{rseuclidian1,rseuclidian2,rseuclidian3,fernandez,
rey,osha,beni} and targets search in discrete lattice 
environments~\cite{santos}; and to avoid the extinction edge 
in scenarios of low availability of energetic 
resources~\cite{faustino}.
%

There is a much less studied class of problems known as
deterministic walks \cite{lima,deterministic,bunimovich}.  
As in the usual stochastic case, they describe the movement of 
a walker in a certain medium, which can or cannot have a random 
character.  
However, the rule of locomotion is always taken from some purely 
deterministic model, rather than from a probability 
distribution~\cite{bunimovich}.
Deterministic walks usually present the technical difficulties 
common to nonlinear dynamical systems~\cite{lima} and can 
give rise to superdiffusive processes~\cite{bunimovich}.  
In fact, they belong to a new class of models known as local 
optimization problems, such as the traveling tourist~\cite{lima}.  
In contrast to the previously discussed examples of purely random 
walks, it seems that for deterministic walks there are no general 
guidelines indicating when the evolution would generate power-law 
distributions for the dynamical variables.

In the present contribution we study the previous general 
questions for a specific deterministic walk.  
We revisit a recently proposed model~\cite{boyer}, in which the 
walker moves in straight lines from site to site, following 
a ``go to the closest target site'' rule.
The sites are randomly distributed in a 2D region.
As already pointed out in~\cite{boyer}, for certain very 
particular parameter conditions, this type of dynamics
surprisingly exhibits power law distribution of step lengths.  
Here we reveal the mechanisms leading to such behavior, not 
analyzed in~\cite{boyer}, showing that the crossover is due 
to a trapping effect associated to particular spatial 
configurations of the landscape.  
The onset of this phenomenon resembles a critical point in 
thermodynamics, even though there is no real phase transition 
in the system.  

The paper is organized as follows.  
In Sec. II we propose the model.  
Simulations are presented in Sec. III.  
In Sec. IV we discuss and interpret our findings.  
Final remarks and conclusion are drawn in Sec. V.

\section{The model}

\begin{figure}
\vspace{1cm}
\centerline{\psfig{figure=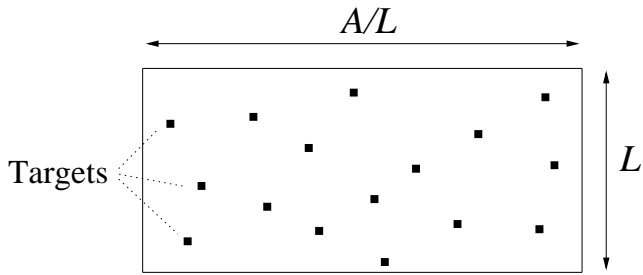,width=8.5cm}}
\caption{
Schematics of the search space where the small squares
represent the randomly distributed target sites.
}
\medskip
\label{ambiente}
\end{figure}

\begin{figure}
\centerline{\psfig{figure=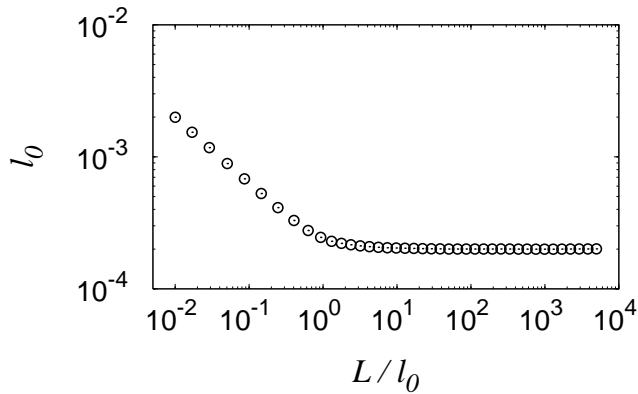,width=8.5cm}}
\caption{
$\ell_0$ as function of $L/\ell_0$.
For $L / \ell_0 \gg 1$ (2D regime), $\ell_0$ is the constant
$1/\sqrt{N}$, whereas for $L/ \ell_0 \ll 1$ (1D),
$\ell_0$ goes as $1/(NL)$.
The crossover takes place for $L/\ell_0$ around unity.
}
\label{flambda}
\medskip
\end{figure}

\begin{figure}
\centerline{\psfig{figure=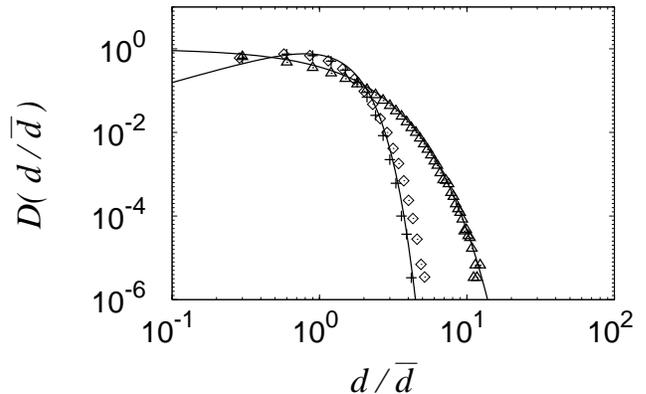,width=8.5cm}}
\caption{
The numerically calculated distributions of distances
between closest neighbor sites for three values of 
$L/\ell_0$ (see main text), corresponding to the:
2D (cross), 1D (triangle) and crossover region (diamond)
cases.
The respective analytical distributions fitting 2D and 1D
are plotted as solid lines. 
Note that the intermediate (crossover) is close to the 2D 
case.
}
\label{finitial}
\medskip
\end{figure}

We consider a deterministic walk model that was originally 
presented in Ref.~\cite{boyer} to describe the locomotion of 
spider monkeys during foraging~\cite{fernandez}.
We define a rectangular region of area $A$ 
and length $L_1 = L$ and $L_2 = A/L$ along the vertical 
($y$-axis) and horizontal ($x$-axis) directions, respectively.
Within this domain, a total of $N$ point targets are 
initially distributed at random.
The configuration of the search region is schematically
represented in Fig. 1.
In all the simulations we set $N = 2.5 \times 10^7$ and $A=1$.
The dynamics is given by the two following simple rules:
\begin{itemize}
\item 
Once at a certain target site, the walker moves straight
to the closest available site.
\item 
The walker does not come back to any previously visited site
-- the search is destructive, i.e., the total number of sites
decreases as they are found along the walk.
\end{itemize}

Let us define the characteristic length 
\begin{equation}
\ell_0=2\bar{d}
\end{equation}
with
\begin{equation} 
\overline{d} = \frac{1}{N}\sum_n d_n,
\end{equation}
where $d_n$ is the distance between the target $n$ and its closest 
neighbor. 
As $L$ can be taken in the interval $[0,1]$, we have two limiting 
cases.
When $L = {\cal O}(1)$, the process takes place in a 2D space and  
$\ell_0 = \sqrt{1/N}$.
On the other hand, as $L \rightarrow 0$ the domain is 1D
and $\ell_0 = 1/(L N)$.
The crossover between these two regimes is found by varying $L$.
Fig. 2 displays the numerically calculated $\ell_0$ as a function 
of $L/\ell_0$.
The two limiting behaviors are clearly seen and separated by a 
crossover emerging around $L/\ell_0 \approx 1$.
In the following, we will use $L/\ell_0$ as the main parameter
of the model.

In Fig. 3 we display the distribution $D(d/\overline{d})$ of 
the separation distances $d_n$ for the three situations:
the 2D limit with $L/\ell_0 \approx 4978.56$
($L = 1$ and $\ell_0 = 2.00861 \times 10^{-4}$),
the 1D-limit with $L/\ell_0 \approx 9.99474 \times 10^{-3}$
($L = 2 \times 10^{-5}$ and $\ell_0 = 2.00105 \times 10^{-3}$),
and in the crossover region with $L/\ell_0 \approx 4.21598$
($L = 8.82 \times 10^{-4}$ and $\ell_0 = 2.09204 \times 10^{-4}$). 
The expected distributions, 
Poisson $D(d/\bar{d}) = \exp[-d/\bar{d}]$ and the standard 
Weibull (i.e., a weighted Gaussian)
$D(d/\bar{d}) = (\pi/2) \, d/\bar{d} \, 
\exp[-\pi d^2/(4 \bar{d}^2)]$, respectively in the
1D and 2D cases, are recovered.


\section{Results}

\begin{figure}
\centerline{\psfig{figure=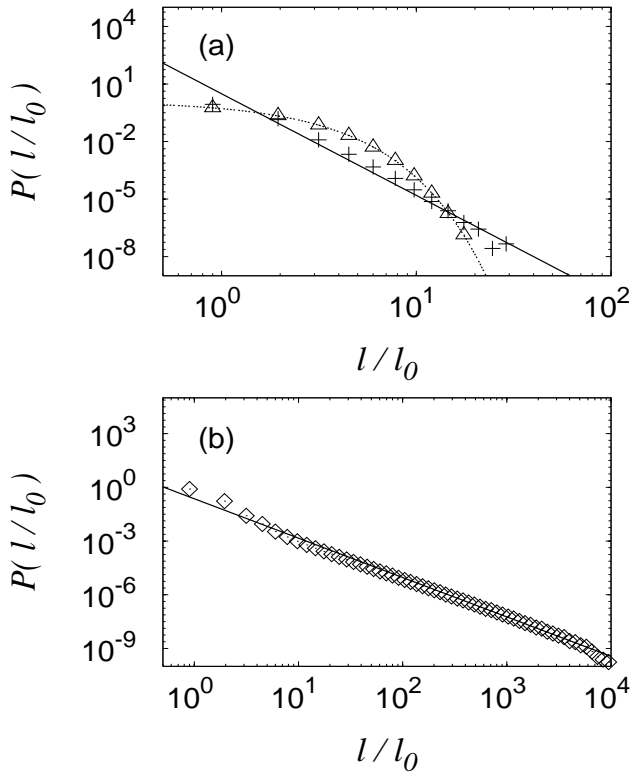,width=8.5cm}}
\caption{
Normalized step length distributions for the same parameters 
as in Fig. 3.
(a) The triangles (cross) represents the 1D (2D) limit.
The curves show the fits
$1.3 \, \exp[- 0.92 \, \ell/\ell_0]$ (dashed) and 
$3.1 \, (\ell/\ell_0)^{-5.3}$ (continuous).
(b) The intermediate case (diamond).
Here the fit is $0.23 \, (\ell/\ell_0)^{-2.2}$.
}
\label{floglevy}
\medskip
\end{figure}

\begin{figure}
\centerline{\psfig{figure=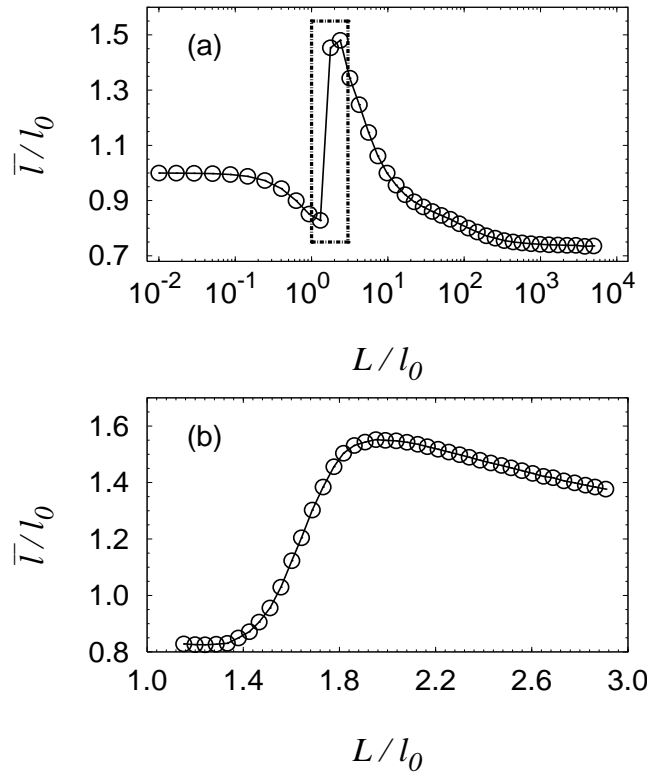,width=8.5cm}}
\caption{
(a) The numerical average step length $\overline{\ell}$ 
(in units of $\ell_0$) taken by the walker during the search 
as function of $L/\ell_0$.
The region where $\overline{\ell}$ presents a peak, indicated 
by a rectangle, is shown in detail in (b).
The continuous curves are just guides for the eye.
}
\label{fmean}
\medskip
\end{figure}

We simulate the walks by iterating our two rules for 
various values of $L/\ell_0$.
Each simulation runs until the walker visits $10^5$ 
targets, out of the initial $N = 2.5 \times 10^7$. 
The curves are obtained by averaging over $10^3$ runs. 
At $t=0$, the walker is located on a site in the vicinity 
of the center of the searching environment.

A first important quantity is the distribution $P(\ell/\ell_0)$ 
of the reduced distance $\ell/\ell_0$ traveled by the walker 
between two consecutive targets sites.  
In Fig. 4(a) we show the results corresponding to the
1D and 2D cases of Fig. 3.
For 1D, the curve follows the expected exponential Poisson 
distribution.
For 2D, $P(l/\ell_0)$ differs markedly from the standard 
Weibull (i.e., a weighted Gaussian) distribution of nearest 
distances $D(d/\overline{d})$.  
The curve is broader, but can be well fitted by a rapidly  
decaying inverse power-law with exponent close to $5.3$,
whose general behavior is actually Gaussian, driven
by the Central Limit Theorem with converging second
moment~\cite{shlesinger1}.
However, for the example in the crossover region, 
$L/\ell_0 \approx 4.21598$, $P$ clearly exhibits a very 
long tail, as shown in Fig. 4 (b).  
There is a small but non-negligible probability of long walks.  
In this case, we find numerically that $P \sim (\ell/\ell_0)^{-\mu}$ 
with $\mu \approx 2.2$ ($\mu \approx 2.15$ by considering only the 
interval $10 < \ell/\ell_0 < 10^4$).  
Thus, the distribution has a power-law behavior with a diverging 
second moment, similar to L\'evy processes.

A second relevant quantity is the reduced average step length 
$\overline{\ell}/\ell_0$, displayed in Fig. 5 (a) as a 
function of $L/\ell_0$.
As shown in more detail in Fig. 5 (b), we have a peak for
$\overline{\ell}/\ell_0$ in the crossover region, with the 
maximum corresponding to $L/\ell_0 \approx 1.99 \approx 2$.

From the Fig. 5 we are lead to think that the step lengths 
in the crossover region are indeed larger than those for 
the 1D and 2D limits.
In fact, we find this is true within the interval 
$2 < L/l_0 < 30$, where the distribution $P(\ell/\ell_0$) can 
fairly be written as $(\ell/\ell_0)^{-\mu}$ with 
$2 < \mu < 3$.
For some values of $L/l_0$, we list in Table I the 
corresponding power-law exponents $\mu$.
Some step lengths distributions are shown in Fig. 6.
For comparison we also plot the example of Fig. 4 (b).

\begin{figure}
\centerline{\psfig{figure=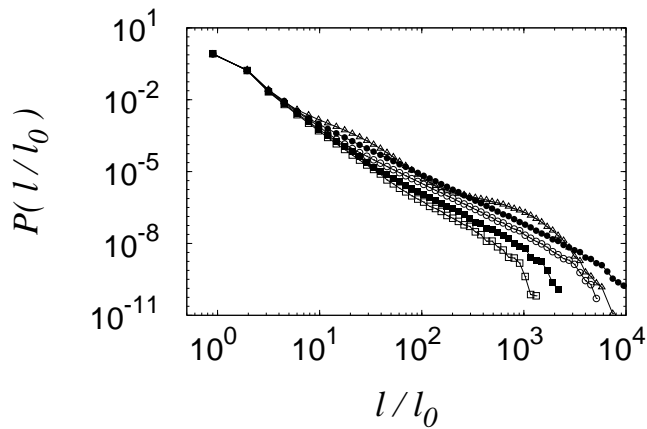,width=8.5cm}}
\caption{The distribution $P(\ell/\ell_0)$ fitted as
$(\ell/\ell_0)^{-\mu}$.
The parameters are:
$L/\ell_0 = 2.37957$ and $\mu = 2.3155$
(open triangle);
$L/\ell_0 = 4.21598$ and $\mu = 2.22267$
(full circle);
$L/\ell_0 = 9.74390$ and $\mu = 2.40229$
(open circle);
$L/\ell_0 = 22.1615$ and $\mu = 2.65706$
(full square); and
$L/\ell_0 = 38.2288$ and $\mu = 2.96385$
(open square).}
\medskip
\end{figure}

\begin{table}
\label{tableI} 
\caption{For some values of $2 < L/\ell_0 < 30$,
the fitted $\mu$'s for
$P(\ell/\ell_0)$ written as $(\ell/\ell_0)^{-\mu}$.}
\begin{tabular}{c c}
\hline
\hline
\ $L/l_0$ \ \ \ \  &
\ \ \ \ \ \ $\mu$  \ \ 
\\ \hline
2.37957  \ \ \  &  \ \ \  2.31550
\\
3.17214  \ \ \  &  \ \ \  2.28575
\\ 
4.21598	 \ \ \  &  \ \ \  2.22267 
\\
5.58647	 \ \ \  &  \ \ \  2.28323
\\
7.38735	 \ \ \  &  \ \ \  2.33486
\\
9.74390	 \ \ \  &  \ \ \  2.40229
\\
12.8287	 \ \ \  &  \ \ \  2.48336
\\
16.8655	 \ \ \  &  \ \ \  2.52824
\\
22.1615	 \ \ \  &  \ \ \  2.65706
\\
29.1083	 \ \ \  &  \ \ \  2.78089
\\
\hline \hline
\end{tabular}
\end{table}

\section{Discussion}

\begin{figure}
\centerline{\psfig{figure=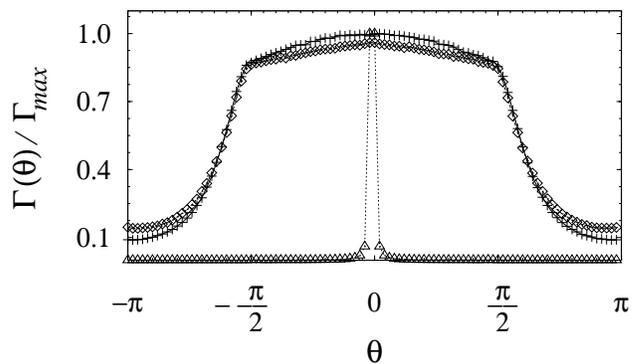,width=8.5cm}}
\caption{Angular distribution of the turning angles 
between consecutive steps.  
The triangles (1D), crosses (2D) and diamonds 
(crossover region) correspond to the same cases 
of Fig. 4.
The dotted curve (for the triangles) is just a 
guide for the eye.}
\label{angle}
\medskip
\end{figure}

\begin{figure}
\vspace{1cm}
\centerline{\psfig{figure=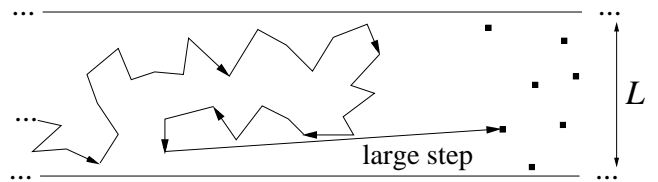,width=8.5cm}}
\caption{In the crossover regime, the search space is a narrow strip.  
In such case, a walker moving in average to the right may perform 
many steps in the opposite direction, returning near the rightmost 
visited site through a very long step.}
\medskip
\end{figure}

To understand the above results we turn to the dynamics of the
deterministic search process. 
In the 1D limit ($L \rightarrow 0$) the random walker tends 
to follow an almost straight line, with only a few changes 
in direction, mostly occurring during the first steps. 
On the other hand, the 2D limit ($L \rightarrow 1$) is 
characterized by a much larger available space in both directions.
Although the destruction of previously visited sites makes 
the walker tend to move forwards with higher probability, 
there is a finite fraction of large turning angles along the walk.
To quantify these features we present in Fig. 7 a normalized 
plot of the angular distribution $\Gamma(\theta)$ of angles 
between two consecutive steps corresponding to the examples
of Fig. 4. 
In the 1D limit, the distribution is very peaked 
at small angle values, indicating that the walker rarely deviates 
from a certain direction (either left or right, defined shortly after 
a few initial steps). 
In the 2D limit there is a bias toward the forward direction, 
nevertheless larger turning angles are also likely to happen.

It is, however, interesting to notice that Fig.~7 alone is not
sufficient to explain our findings, since the 2D and crossover 
cases present similar $\Gamma(\theta)$.  
As the search space shrinks in one direction, we pass through a 
crossover region from 2D to 1D.
The singular behavior in this intermediate regime can be 
explained in terms of a recurrent feature observed in our 
simulations (see also~\cite{boyer}).  
In this case the walker moves in average towards a given direction, 
say to the right, in a narrow strip-like space.  
However, sometimes it turns to an anti-parallel path to visit
sites left behind.  
After some time, the walker ends up in a region depleted of 
targets, which may be far away from the rightmost point reached 
by the trajectory.  
To return to the region rich in unvisited targets located to 
the right, the walker then needs to make a long jump, as depicted 
in Fig. 8. 
Such mechanism is illustrated in Fig. 9, that shows a space-time 
graph of a simulated trajectory in the three different regimes 
of Fig. 4.

\begin{figure}
\vspace{1cm}
\centerline{\psfig{figure=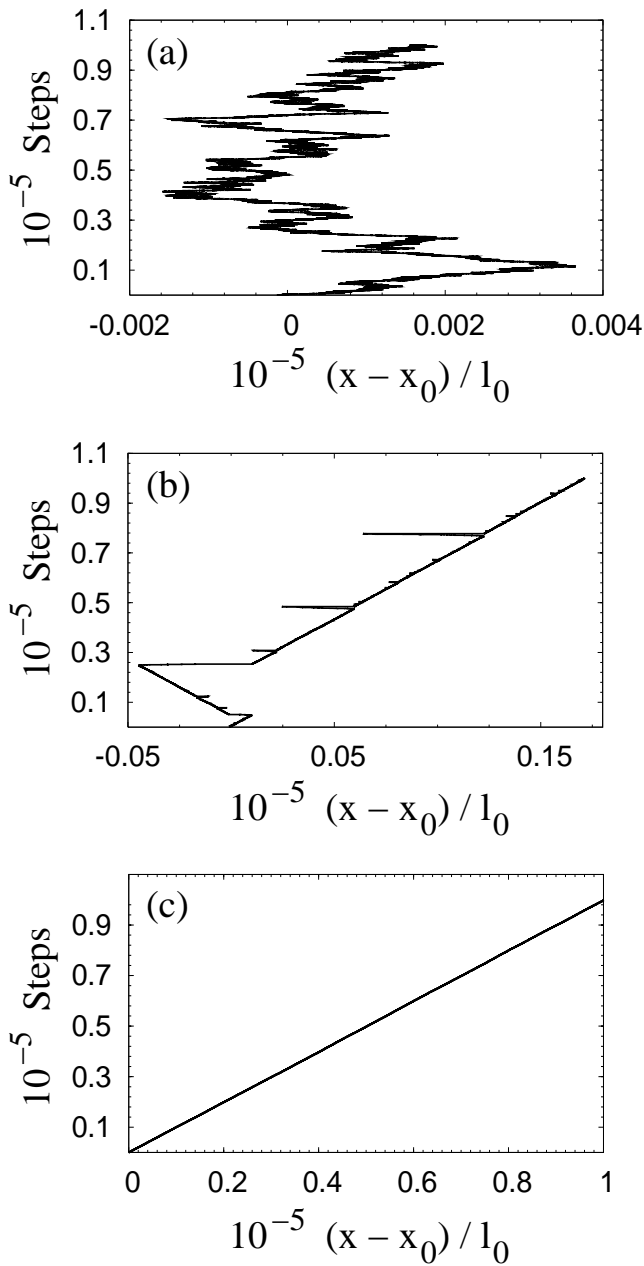,width=8.5cm}}
\caption{Number of steps vs. the horizontal projection of the walker
position, in the 2D (a), crossover region (b), and 1D (c) cases of
Fig. 4. Notice the crucial ultra-long steps at the crossover geometry
(b).}
\medskip
\end{figure}

This dynamical process is particularly sensitive to the 
distance between the two horizontal borders or equivalently, 
to the values of $L/\ell_0$.
In fact, for very small $L/\ell_0$ (1D) there are no anti-parallel
paths, whereas for large $L/\ell_0$ (2D) the extra vertical direction
often provides closer sites than those reached by big jumps across 
depleted regions.
None of these two aspects, the directional bias in the 1D case 
and the extra dimension providing many "escape" paths, 
are present in the crossover region.
Note that the broad power-law distributions for the length of 
the steps $\ell$ (Fig. 4 (b)) are observed when the probability 
of large turning angles is high.
In the crossover region it is higher than in the 2D regime, 
as shown in Fig. 7.


The above scenario is confirmed by analyzing two quantities
related to the dynamics of the deterministic search process.
First, we calculate the normalized drift velocity along the 
horizontal direction $x$, defined as 
$\left<|x-x_0|/n\right>/\ell_0$, where $x_0$ is the starting 
coordinate and $x$ is the coordinate at step $n$.
We show in Fig. 10 (a) the drift velocity as a function of 
$L/\ell_0$. 
As expected, it vanishes in the 2D limit.
Worthwhile noticing, however, is the behavior of the curve in 
the crossover region, Fig. 10 (b). 
In particular, it presents a local minimum around 
$L/\ell_0 = 1.7094$.
Furthermore, the first local maximum after this minimum 
is at $L/\ell_0 = 2.0368 \approx 2$,
the same position for the maximum of $\overline{\ell}/\ell_0$
seen in Fig. 5 (b).

\begin{figure}
\centerline{\psfig{figure=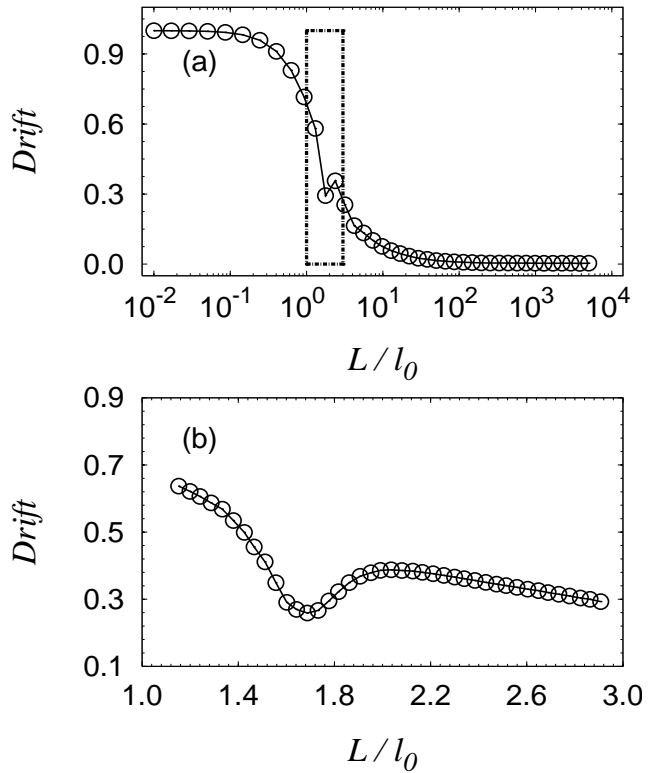,width=8.5cm}}
\caption{
(a) The numerical drift velocity along $x$,
$\left<|x-x_0|/n\right>/\ell_0$, as a function of $L/\ell_0$.
The rectangle indicates part of the crossover region where 
there is an inflection point for the drift.
(b) Blow up of the marked region in (a), showing a local
minimum around $L/\ell_0 = 1.7094$.
The continuous curves are just guides for the eye.
}
\label{fdrift}
\medskip
\end{figure}

A second relevant quantity is the fraction of visited targets
along the walker trajectory, defined by
\begin{equation}
\chi = \frac{M_{\mbox{\scriptsize vis.}}}{M_0}.
\end{equation}
Here, $M_{\mbox{\scriptsize vis.}}$ is the average number of 
visited targets in the area searched by the forager and $M_0$ 
is the total number of initial targets in that area. 
The searched area is defined by the region
$[x_{\mbox{\scriptsize max}} - x_{\mbox{\scriptsize min}}] 
\times 
[y_{\mbox{\scriptsize max}} - y_{\mbox{\scriptsize min}}]$,
where the \lq\lq min" and \lq\lq max" subscripts 
stand for the minimum and maximum values of the coordinates 
reached by the walker during a full run.
Thus, $\chi$ represents a search efficiency of the walker. 
For the 1D regime, all the targets are found along
the way, so $\chi = 1$. On the other hand, at 2D regime, $\chi$
takes a constant small value. The crossover from one limit to the
other, as a function of $L/\ell_0$, is shown in Fig. 11.
Again we observe a local minimum in the crossover region,
as seen in the inset of Fig. 11.

\begin{figure}
\centerline{\psfig{figure=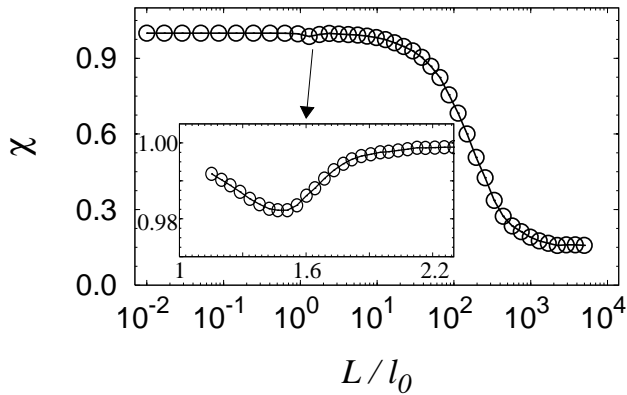,width=8.5cm}}
\caption{
The numerical fraction of visited targets, $\chi$, as a 
function of $L/\ell_0$.
The inset shows a local minimum in the crossover region.
The continuous curves are just guides for the eye.
}
\label{fchi}
\medskip
\end{figure}

\section{Final Remarks and Conclusion}

Motivated by the results of a previous study \cite{boyer}, we 
have investigated in detail a deterministic walk model where the
destructive search environment can be changed from a 2D to a 1D
geometry by tuning a single control parameter, namely, 
$L/\ell_0$.  
The movement of the walker is driven by the ``go to closest 
target'' rule. 
Naively, the model should lead to a Poisson process, since the
initial distribution of target sites (which are destroyed once
visited) is random. 
We actually find that in both the 1D and 2D limiting cases, 
the step lengths distribution have finite variance.
However, for some intermediate values of $L/\ell_0$, a 
non-trivial dynamical process with infinite variance takes place, 
combining a large number of relatively small steps with rare long 
steps.  
It gives rise to a \lq\lq L\'evy-like" step length distribution, 
in which the probability of large steps is enhanced by 
characteristic exponents in the range $2 < \mu < 3$, e.g.,
$\mu \approx 2.2$ for the case in Fig. 4 (b).  
Furthermore, for values of the control parameter in this 
crossover region, we observe changes in the search efficiency 
and drift velocity.  
Such findings are interesting since they show that power-law 
distributions can also result from a simple short-range dynamics 
combined with a geometrical constraint.

Finally, we comment on interesting similarities between our model and
other related problems.  
First, a random foraging model based
on L\'evy strategies for target sites with a recovery (or regeneration
or refractory) delay time \cite{rseuclidian2,rseuclidian3} has been
recently studied.  Once visited, a target site become available for a
future visit only after a finite number of steps (time delay $\tau$).
It has, as limiting regimes, the destructive ($\tau \rightarrow
\infty$) and non-destructive ($\tau \rightarrow 0$) random searches.
Differently from the present case, the walker may either finish a
given step with no target found or truncate its flight if a site is
found along its way.  In this foraging problem the most efficient
destructive (non-destructive) searches requires $\mu \approx 1$ ($\mu
\approx 2$).  
So, the parameter $\tau$ makes the crossover from one to
other limit.  
Thus, it seems that the presence of boundaries in the
deterministic walk and the time delay in the foraging random search
problem play similar roles in the sense that they determine the
characteristic exponent for the distribution of the step lengths of
the respective walkers, governing then the type of dynamics.

Second, the foregoing results also suggest an analogy between our 
deterministic walker model and thermodynamic systems and
phase transitions.
The observed motion in the 2D limit is relatively isotropic,
(of course with some bias due to ``a back step'' depletion),
whereas the motion in 1D breaks completely this isotropy.  
Similarly, the behavior of the velocity is ergodic in the 2-D 
limit but non-ergodic in 1D limit.  
Let us now consider more carefully the crossover.  
Exactly at the point in between the two behaviors, we expect the 
velocity to be marginally ergodic, such that the average behavior 
of the velocity inversion becomes log-periodic rather than periodic 
(2D limit) or nonperiodic (1D limit).  
Log-periodic velocity inversions~\cite{cressoni} can represent 
the border between superdiffusive and diffusive regimes.  
Moreover, this logarithmic behavior implies that mean values 
of $\ell(t)/\ell_0$ will scale geometrically with time $t$, such 
that the mean value of $\ell$ diverges. 
In other words, the larger the system size (or simulation time), 
the larger the mean value of $\ell$.  
Except for the memory or correlation effects, this is the same 
kind of behavior we find in L\'evy walks.  
The maximum superdiffusion for L\'evy walks occurs when the 
first moment of the mean step size diverges, which corresponds to
an inverse square distribution of $\ell$. 
Considered from this point of view, the results in Figs. 4 (b) 
and 5 make qualitative sense.

Thus, although the present study indicates that the system 
is going through a crossover between two different limits, from 
the above discussion we cannot completely rule out the possibility 
of a dynamical phase transition.
This issue is presently being investigated and will be reported 
in the due course.


\section*{Acknowledgments}

Luz, Raposo, Santos and Viswanathan acknowledge CNPq, 
CNPq/Edital Universal, FACEPE, FAPEAL, 
Funda\c c\~ao Arauc\'aria and Finep/CT-Infra for research 
grants and also are greatfull to Prof. Carlos Carvalho 
for very helpful computational hints.
Miramontes and Boyer acknowledge UNAM-DGAPA grant IN118306
and Conacyt grant 40867-F for financial support.
Finally, Luz thanks the Physics Institute at UNAM for 
the kind hospitality during stays in which this work has 
been developed.

\end{document}